\documentclass{eas}
\usepackage{graphicx}
\usepackage{journals}
\usepackage{natbib}
\begin{document}

\TitreGlobal{Mass Profiles and Shapes of Cosmological Structures}

\title{Kinematical and Dynamical Modeling of Elliptical Galaxies}
\author{Mamon, G. A.}\address{IAP, 98 bis Bd Arago, F-75014 Paris, FRANCE}
\author{{\L}okas, E.}\address{Copernicus Astronomical Center, 
Bartycka 18, PL-00-716 Warsaw, POLAND}
\author{Dekel, A.}\address{Racah Institute of Physics, Hebrew University, Jerusalem, ISRAEL}
\author{Stoehr, F.$^1$}
\author{Cox, T. J.}\address{Harvard-Smithsonian CfA, 60 Garden Street,
Cambridge MA 02138, USA} 
\runningtitle{Kinematical \& dynamical modeling of ellipticals}
\setcounter{page}{139}
\index{Mamon, G. A.}
\index{{\L}okas, E.}
\index{Dekel, A.}
\index{Stoehr, F.}

%
\begin{abstract}
Elements of kinematical and dynamical modeling of elliptical galaxies are
presented.
In projection, NFW models resemble S\'ersic models, but with a very narrow
 range of shapes ($m=3\pm1$).
The total density profile of ellipticals cannot be NFW-like because
 the predicted local $M/L$ and aperture velocity dispersion within an
effective radius ($R_e$) are much lower than
 observed. Stars must then dominate ellipticals
out to a few $R_e$.
Fitting an NFW model to the total density profile of S\'ersic+NFW (stars+dark
 matter [DM]) ellipticals results in very high concentration parameters, as
found 
 by X-ray observers.
Kinematical modeling of ellipticals assuming an isotropic NFW DM model
underestimates $M/L$ at the virial radius by a factor of 1.6 to 2.4, because
dissipationless $\Lambda$CDM halos have slightly different density profiles
and slightly radial velocity anisotropy. 
In $N$-body+gas simulations of ellipticals as merger remnants of
spirals embedded in DM halos,
the slope of the DM density profile is steeper
when the initial spiral galaxies are gas-rich.
The Hansen \& Moore (2006)
relation between anisotropy and the slope of the density profile breaks
 down for gas and DM, but the stars follow an analogous relation with
 slightly less radial anisotropies for a given density slope.
Using kurtosis ($h_4$) to infer anisotropy in ellipticals is dangerous, as
 $h_4$ is also sensitive to small levels of rotation.
The stationary Jeans equation provides accurate masses out to $8\,R_e$.
The discrepancy between the modeling of Romanowsky et al. (2003),
indicating a dearth of DM in ellipticals,
and the simulations analyzed by Dekel et al. (2005), which
match the
spectroscopic observations of ellipticals,
 is partly due to 
radial anisotropy and to observing oblate ellipticals face-on.
However,
one of the 15 solutions 
to the orbit modeling of Romanowsky et al. is found to have an amount and
concentration of
DM
consistent with 
 $\Lambda$CDM predictions.
\end{abstract}
\maketitle
%
\section{Introduction}
The quantity of dark matter lying in the outskirts of luminous 
elliptical galaxies is
a hotly debated topic (see Romanowsky, Napolitano, Stoehr, in these
proceedings). 
There is a wide consensus that, given their flat rotation curves,
spiral galaxies must be embedded within dark
matter halos, unless one resorts to modifying
physics (e.g. MOND, see McGaugh in these proceedings).
Moreover, dissipationless cosmological $N$-body simulations lead to
structures, whose halos represent most (Hayashi et el. 2004) if not all
\citep{Stoehr06} spiral galaxies.

If elliptical galaxies originate from major mergers of spiral galaxies 
\citep{Toomre77,Mamon92,BCF96,SWTK01},
then
they too should possess dark matter halos.
However, inferring the presence of dark matter halos in ellipticals is
difficult, because the velocity dispersions of the 
usual kinematical tracer, stars, can only be measured out to $2\,R_e$
(effective radii, containing half the projected light). Moreover, the mass
distribution depends on the radial variation of the velocity anisotropy, and
one cannot solve for both, unless one assumes no rotation and makes use of
the 4th order moment (kurtosis) of the velocity distribution 
\citep{vdMF93,Gerhard93,LM03}.

Using planetary nebulae (PNe) as tracers of the dark matter at large radii,
\cite{Romanowsky+03}
found low velocity dispersions for their outermost
PNe, which after some simple Jeans modeling and more sophisticated orbit
modeling led them to conclude to a dearth of dark matter in ordinary
elliptical galaxies. This result is not expected in the standard model of
formation of structure in the Universe and of galaxies in particular.

This has led us 
\citep{Dekel+05} to analyze the final outputs of $N$-body
simulations of spirals merging into ellipticals (\citealp{CPJS04}, Stoehr, in
these proceedings). 
\citeauthor{Dekel+05} show that the line-of-sight
velocity dispersions of their simulated merger remnants are as low as the PNe
dispersions measured by
\citeauthor{Romanowsky+03}
thus removing an important thorn
in the standard model of galaxy formation in a $\Lambda$CDM Universe.
One is led to wonder how could
the kinematical analysis of 
\citeauthor{Romanowsky+03}
lead to a lack of dark
matter, while the dynamical analysis of
\citeauthor{Dekel+05} matches the same set of observations with simulations
including normal amounts of dark matter.

To explain this discrepancy, it is useful to focus first on
some important aspects of
kinematical modeling.

\section{Kinematical modeling}

\subsection{Is the total density profile NFW-like?}

\cite{MNLJ05}
have reported that the projected density profiles of the halos
found in dissipationless cosmological $N$-body simulations are well fit by 
\cite{Sersic68}
models, with the shape
parameter $m\approx 3\pm1$.  This result is not surprising, because these
halos are known to be fairly well fit by the 
\citeauthor*{NFW96} (\citeyear{NFW96}, hereafter NFW)
model,
and 
\cite{LM01} had already shown that projected NFW models
resemble S\'ersic models with $m=3\pm1$.

One should not conclude that elliptical galaxies, which are
known to have S\'ersic surface brightness profiles
\citep{CCDO93},
are $M/L=\rm cst$ NFW models, because
\citep{LM01}: 
1) the range of shape parameters is much narrower for the simulated halos ($m
\approx 3\pm1$) than for observed ellipticals ($1<m<6$ or 10);
2) the normalization of a divergent NFW model with a convergent S\'ersic
model leads to absurd total $M/L$ at $R_e$;
3) given the increase of $m$ with luminosity \citep{CCDO93} and the
decrease of halo concentration with mass 
\citep{NFW97,JS00,Bullock+01},
 the trend of increasing best-fit $m$ for increasing
concentration 
\citep{LM01}
implies the unlikely trend that
elliptical galaxy luminosity decreases with increasing halo mass.

\begin{figure}[ht]
\centering
\includegraphics[width=6.1cm]{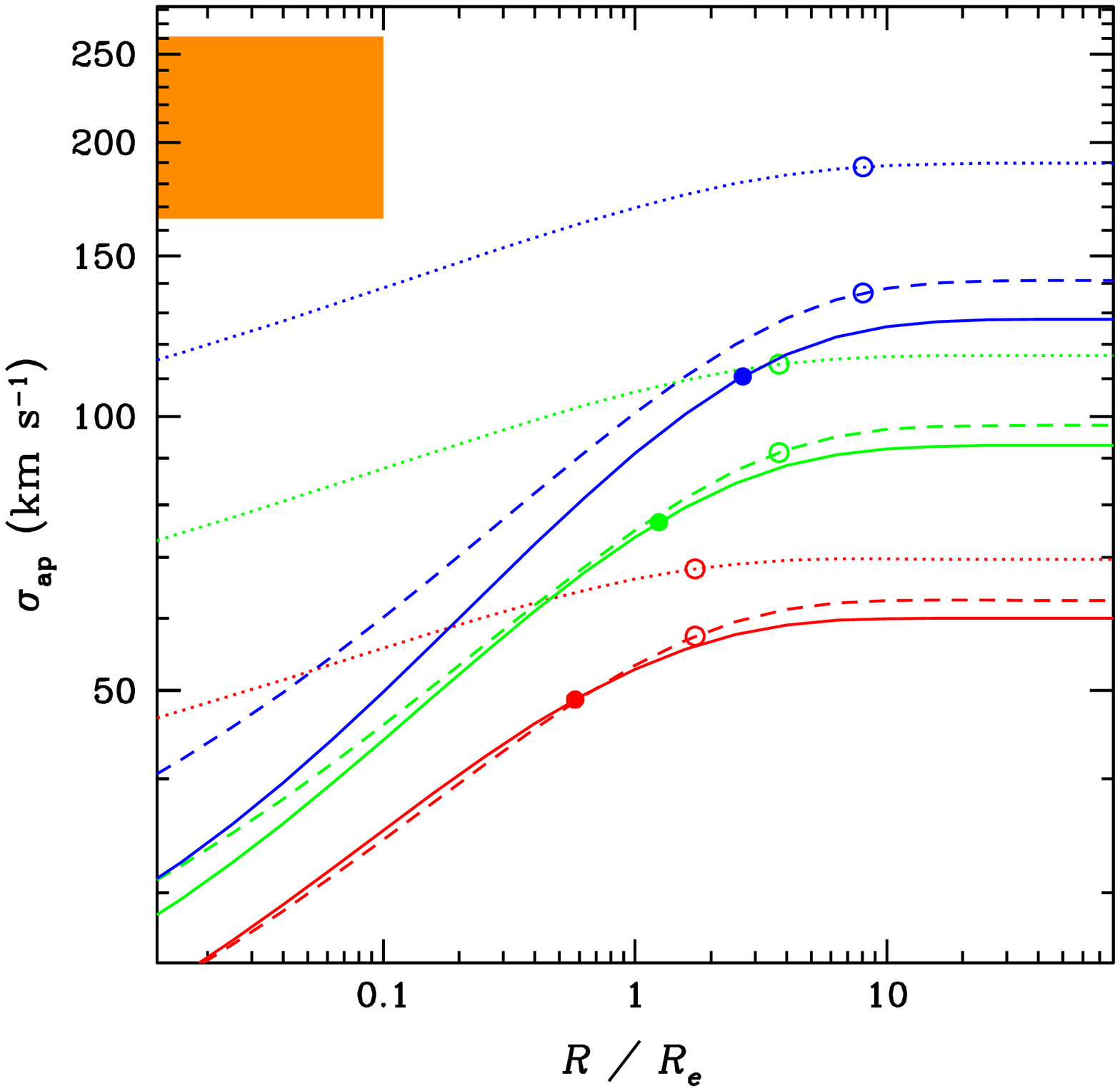}
\includegraphics[width=6.1cm]{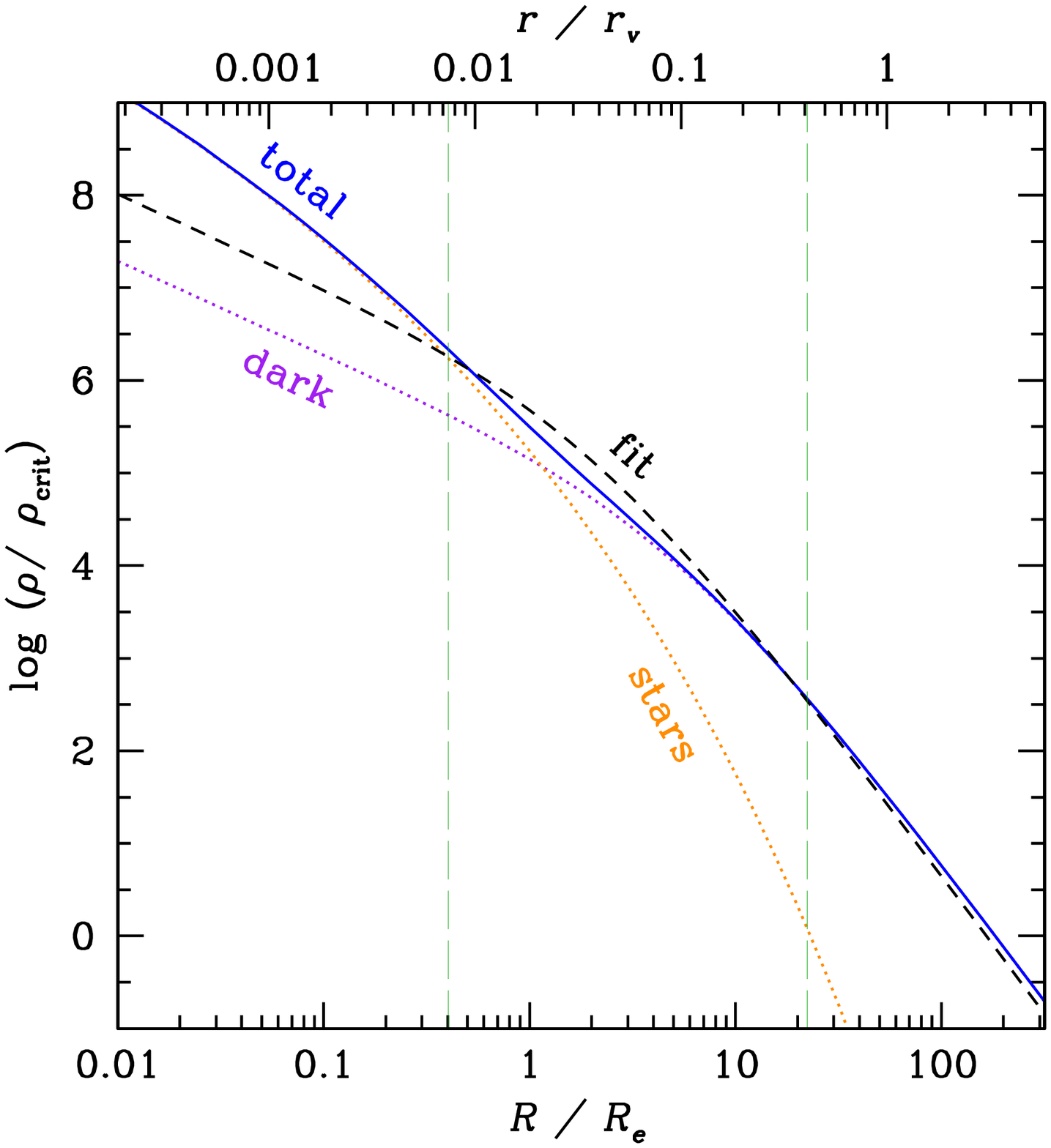}
\caption{\emph{Left}: 
Velocity dispersion averaged over a circular aperture of width $R$ for
models where the total mass density is NFW (\emph{dashed}), Jing-Suto (2005,
their $-3/2$ slope model,\emph{dotted}) and Navarro et al. (2004,
\emph{solid}) curves, for $L_B = L_{B,*} = 1.88\times 
10^{10}L_\odot$ and $M_{\rm vir} / L_B = 39$ 
(\emph{lower curves}), 
390 (\emph{middle curves}), and 
3900 (\emph{upper curves}) $M_\odot/L_\odot$.
The \emph{shaded region} indicates the observational constraints
(Faber-Jackson 1976 relation).
\emph{Right:} density profiles of an elliptical galaxy made of
a S\'ersic stellar and a $c=9$ NFW dark matter component. When fitting an NFW
to the 
total density profile (between the radii of the two \emph{vertical lines}),
a much higher concentration parameter ($c=35$) is found.}
\label{figdenstot}
\end{figure}

Now, it is not clear whether the NFW-like structures found in dissipationless
cosmological simulations represent the dark matter component, i.e. are set by
the dissipationless nature of dark matter, or if they represent the total
matter, i.e. are set by the global gravitational field.
However, if one assumes that the NFW-like models 
represent the \emph{total} mass density of elliptical
galaxies, one 
runs into trouble in two ways 
\citep{ML05a}:
1) as $R$ decreases below $R_e$, the local $M/L$ falls to values
well below the 
$M/L$ of the stellar populations representing ellipticals;
2) the aperture ($R < R_e/10$) velocity dispersions are very much below
the observed values (left plot of Fig.~\ref{figdenstot}) for a given luminosity 
(the \citealp{FJ76} relation).
This occurs with the NFW model, but also with the more recent 3D-S\'ersic
model that 
\citeauthor{Navarro+04} (\citeyear{Navarro+04}, hereafter Nav04)
found to fit better the simulated halos, as well as
the steeper models with inner slope of $-3/2$ proposed by
\cite{FM97,Moore+99} and \cite{JS00}.
%
Given the limited spatial resolution of cosmological simulations
(symbols in Fig.~\ref{figdenstot}), the observed velocity
dispersions could be recovered if the total density profile sharply steepens
(to a slope steeper than $-2$) just at $R_e$, but this seems a little far
fetched.
The low predicted local $M/L$s and
aperture velocity dispersions imply that 
\emph{NFW-like models cannot represent the total matter}.
The simplest
explanation is that these models represent the dark matter
component only and that \emph{a more concentrated stellar component must
dominate 
NFW-like dark
matter
within $R_e$}. 

The right-hand plot of Figure~\ref{figdenstot} shows that \emph{fitting an NFW
model to the total density profile of a system composed of S\'ersic stars and
NFW dark matter yields a high concentration}, as fit by X-ray
observers 
(e.g., \citealp{Sato+00}), and the fit cannot be very good.

\subsection{Weighing the dark matter out to the virial radius}

\begin{figure}[ht]
\centering
\includegraphics[width=8cm]{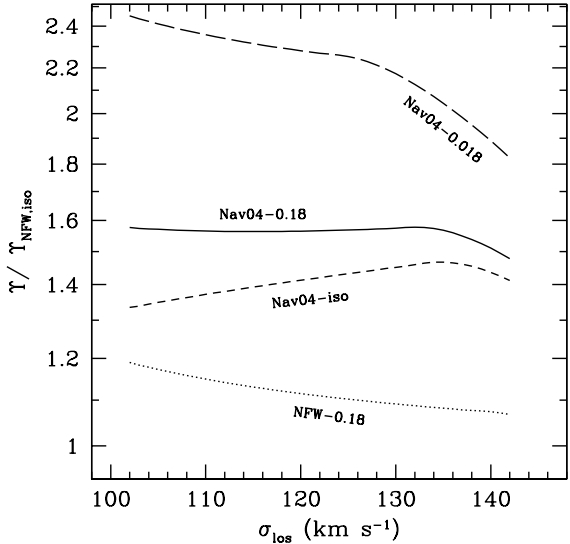}
\caption{$M/L_B$ at the virial radius divided by the 
value obtained assuming an NFW dark
matter model and isotropy, as a function of observed line-of-sight 
velocity dispersion at $5\,R_e$ (adapted from Mamon \& {\L}okas 2005b).
The dark matter model is shown with numbers representing 
the anisotropy radius (where $\beta=1/4$)
in units of $r_v$.
}
\label{upsbias}
\end{figure}

The key physical parameter for the dark matter is its mass within the virial
radius, $r_v$, or equivalently its $M/L$.
Alas, given that observations are very hard to obtain beyond $5\,R_e$
($\approx 0.06\,r_v$), even
with external mass 
tracers such as PNe and globular clusters, 
one requires a strong extrapolation to weigh the dark matter
component out to $r_v$.
\cite{ML05b} find that 
the observed velocity dispersion at $5\,R_e$ is a very weak function of
$M/L$ at $r_v$. Conversely, $M/L$ at $r_v$ is a strong (8th power) function
of the observed $\sigma_v$ at $5\,R_e$.
Figure~\ref{upsbias} illustrates how badly one underestimates $M/L$ at $r_v$ by
assuming an NFW dark matter model instead of the latest 3D-S\'ersic model of
\cite{Navarro+04} and neglecting radial
velocity anisotropy.\nocite{ML05b}
Assuming an isotropic NFW model instead of the slightly radial 
\citeauthor{Navarro+04}
model that matches much better the halos in
cosmological $N$-body simulations the $M/L$ derived at $r_v$ is 60\% too low.
But if the orbits are as radial as found by
\citeauthor{Dekel+05}
in the merger
simulations (see Stoehr, in these proceedings), then $M/L$ is underestimated
by a factor 2.4.

\section{Dynamical modeling}

We have analyzed the end products of $N$-body+gas (SPH) 
simulations of merging equal mass spirals, made of a disk, a bulge, a gas
disk, and a dark matter halo (see Stoehr et al., in these proceedings).
The merger remnants not only show stellar 
surface density profiles that almost perfectly match the observed surface
brightness profiles of ellipticals (an old result), but more interestingly, 
they display the same line-of-sight stellar
velocity dispersion profiles as the stellar and PN velocity dispersion
profiles observed in ellipticals 
(\citealp{Dekel+05}, 
 see Stoehr et al., in
these proceedings).
These simulations highlight how the dark matter and stellar kinematics are
 decoupled. 
The stellar component is found to dominate the dark matter component within
$R_e$
whereas one expects from the dark matter
profiles of cosmological $N$-body simulations and the observed surface
brightness profiles, that the dark matter only dominates the stars at $2\,R_e$
\citep{ML05b}.
Thus the dark matter has readjusted itself in the inner regions where the
stars dominate the gravitational potential.
\begin{figure}[ht]
\centering
\includegraphics[width=7.2cm]{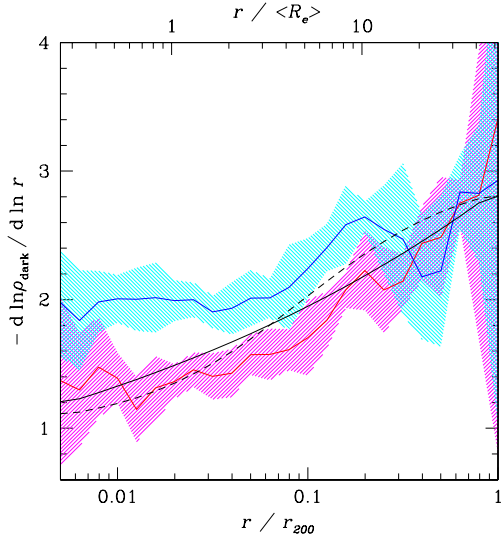}
\caption{Radial profiles of dark matter density slopes from 10 major merger
simulations: \emph{upper curve 
\& region}: gas-rich runs. \emph{lower curve \& region}: gas-poor runs.
\emph{Dashed} and \emph{solid curves} show the predicted slopes for NFW and 
Navarro et al. (2004) dark matter models, respectively, with the
concentration parameters given in Mamon \& Lokas (2005a).}
\label{figetadark}
\end{figure}
Figure~\ref{figetadark} shows that for runs with a low initial 
gas content the dark matter slopes are in good agreement with those seen in
halos of dissipationless $\Lambda$CDM simulations. However, the runs with a
high gas content show higher slopes, as qualitatively expected since the dark
matter should respond to the more predominant inner baryons (see Gnedin,
these proceedings).

\begin{figure}[ht]
\centering
\includegraphics[width=8.5cm]{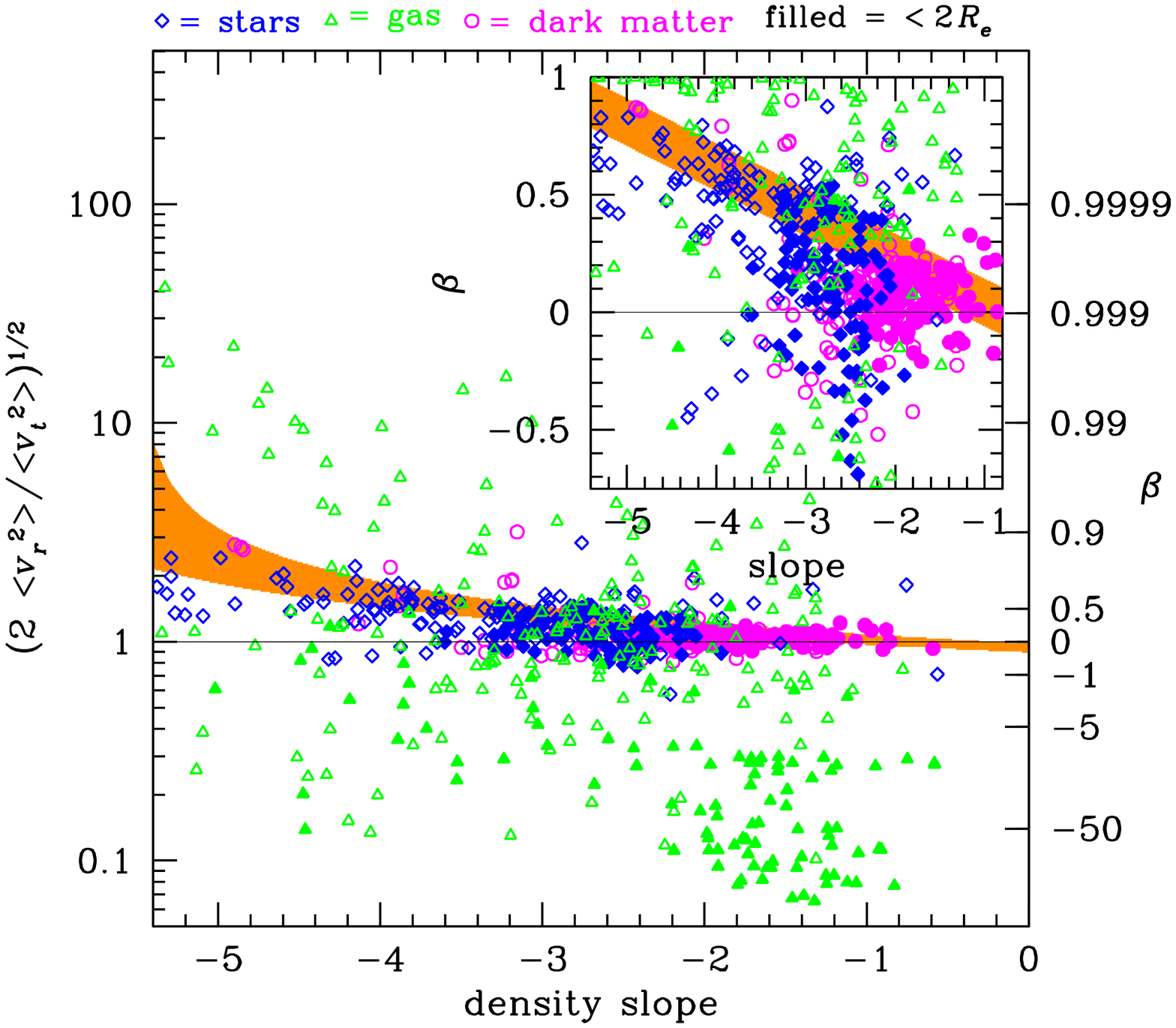}
\caption{Velocity anisotropy vs. logarithmic slope of the
density profile in 10 major meger simulations.
\emph{Circles}, \emph{diamonds} and \emph{triangles} show dark matter, star
and gas particles, 
respectively.
\emph{Filled} and \emph{open symbols} show inner ($R <
2\,R_e$) and outer ($R \geq 2\,R_e$) particles, respectively.
Each symbol represents a spherical shell of one of the 10 runs analyzed.
The shaded regions illustrate $\beta=1-1.15\,(1+\alpha/6)\pm0.1$ ($\alpha$ is
the slope) as found by
Hansen \& Moore (2006) in 
very different dark matter only simulations.
The \emph{thin horizontal lines} represent isotropy.}
\label{figetadarksteen}
\end{figure}
%
\citet{HS06} noticed that in dissipationless simulations,
(both cosmological and non-cosmological), 
the radial profiles of velocity anisotropy and density slope are strongly
correlated, which should, in principle, allow the lifting of
the mass/anisotropy degeneracy (see Sec.~1).
Figure~\ref{figetadarksteen} shows that in the merger
simulations of spiral galaxies made of stars, dark matter and gas, analyzed
by 
\citeauthor{Dekel+05},
each
component obeys its own correlation (or lack thereof) between anisotropy and
density slope. The gas relation shows a rotating inner disk, and 
is very dispersed, presumably because of its
dissipative nature. The dark matter relation is isotropic even in regions of
steep density profiles, in contrast with the
prediction of 
\citeauthor{HS06}.
Interestingly, the stellar relation is close to that of 
\citeauthor{HS06},
but with over double its
dispersion and with $\beta$ roughly 0.15 lower. Presumably the stellar 
kinematics are affected by the high dispersion of the dissipative gas
kinematics from which a few stars are formed.

\begin{figure}[ht]
\centering
\includegraphics[width=10cm]{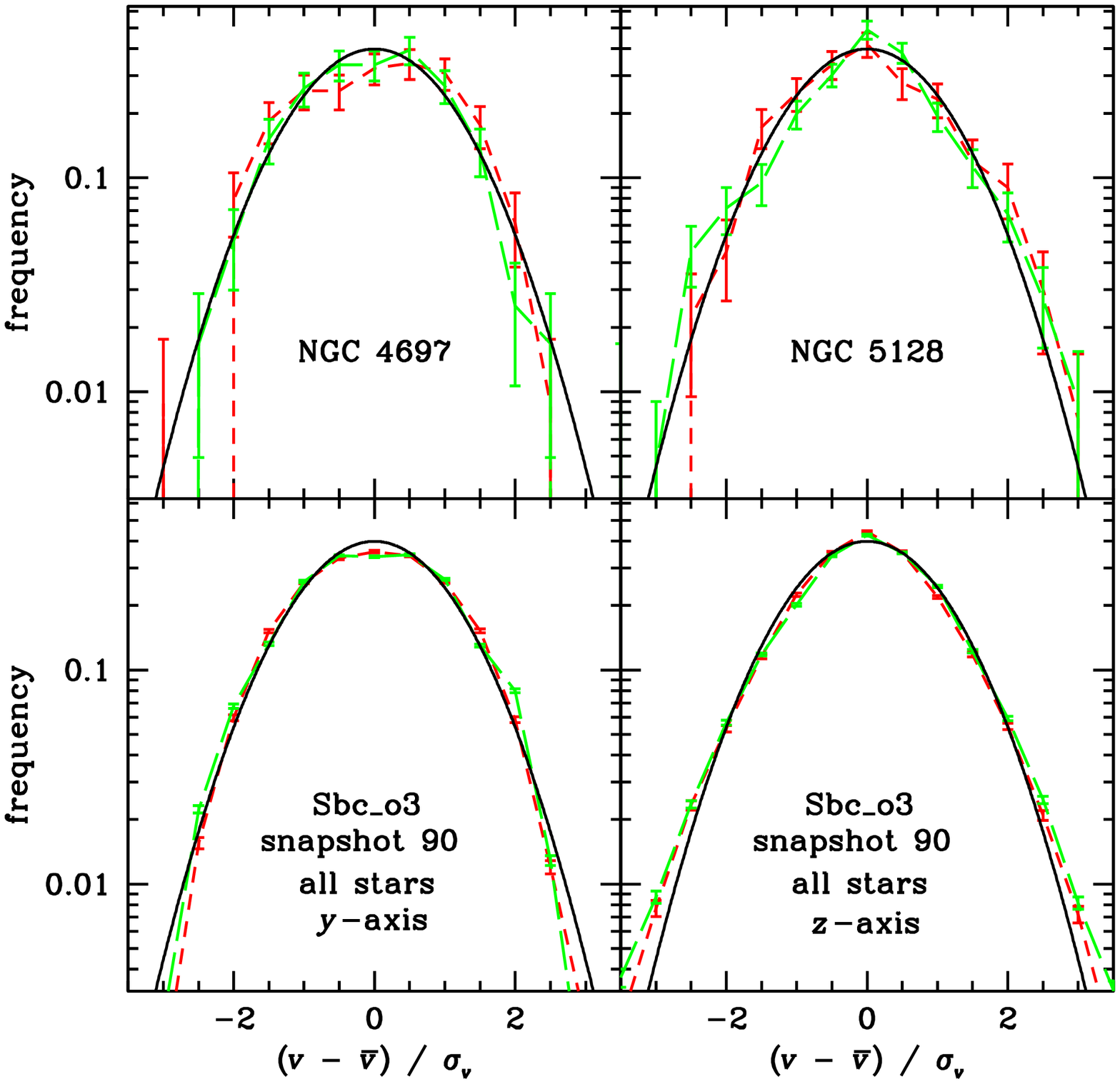}
\caption{Normalized velocity distributions for two observed galaxies
(\emph{top}) and two snapshots of the same simulated merger remnant (stellar
  component)  viewed
perpendicular to (\emph{lower left}) and along (\emph{lower right}) the
angular momentum vector.
\emph{Short-dashed} and \emph{long-dashed curves} represent $0.5\,R_e < R <
1.5\, R_e$ and $R > 1.5\,R_e$, respectively, while the \emph{solid curve} is
the gaussian distribution.}
\label{vdis}
\end{figure}

The simulations also help to understand what drives the 4th order,
gaussian-weighted,  velocity
moment, $h_4$. Figure~\ref{vdis} displays the velocity distribution and $h_4$
values of the PNe observed in two elliptical galaxies as well as the particle
velocity distributions in the same simulation run seen in two orthogonal
projections.
Interestingly, the same snapshot viewed along two orthogonal directions
produces either flat-topped or cuspy velocity distributions. Although the 3D
configuration is radially anisotropic beyond $1.5\,R_e$, the $h_4$ parameter
is either negative (left plot) or positive (right plot). The negative $h_4$ is
caused by a small amount of rotation in the system viewed perpendicular to
the angular momentum axis.

\section{Understanding the low outer velocity dispersions}

\subsection{Lack of equilibrium?}

\begin{figure}[ht]
\centering
\includegraphics[width=8cm]{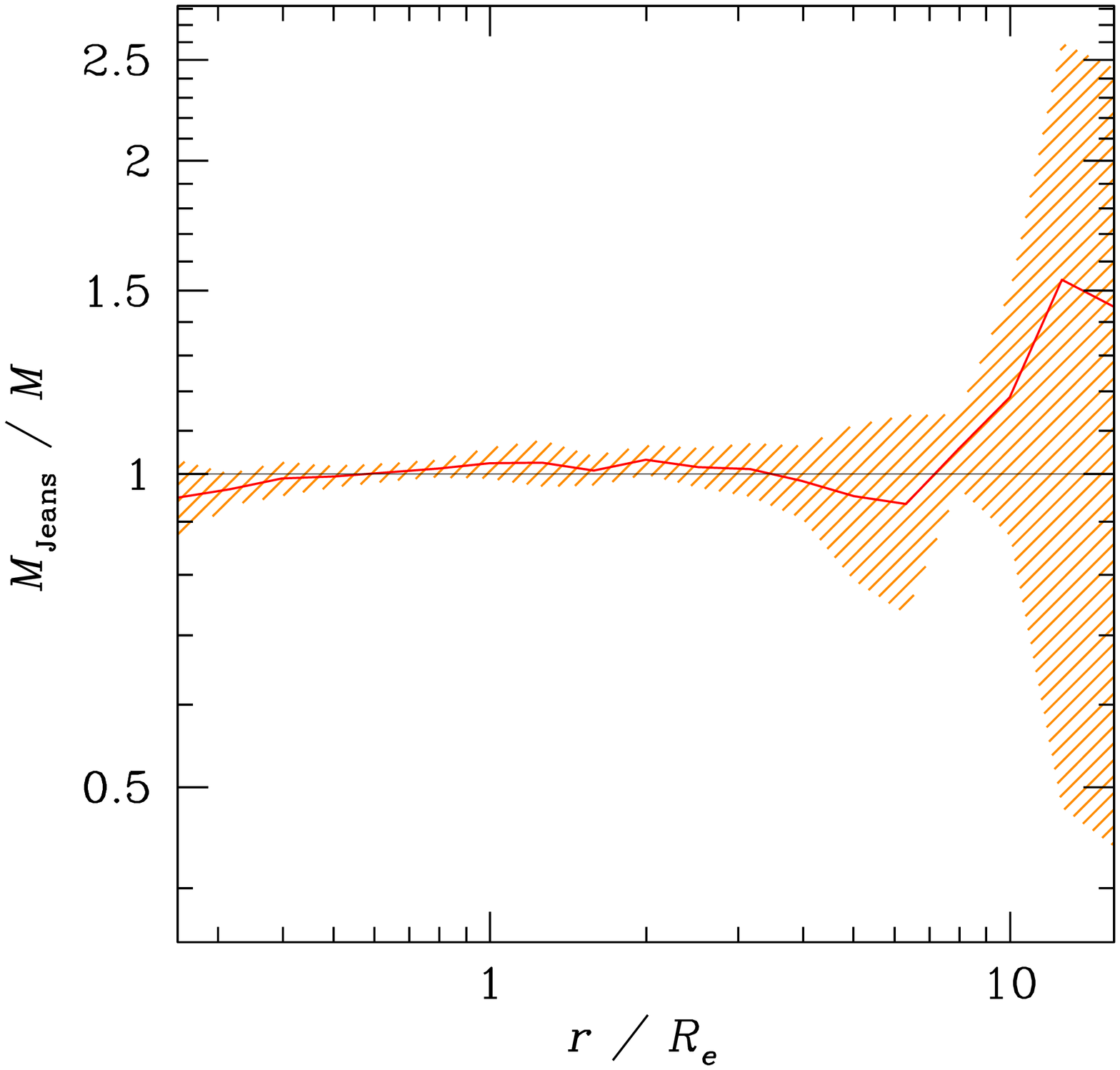}
\caption{Jeans mass (eq.~[\ref{mjeans}])
over true mass in spherical shells for the final times of 10 major merger
simulations.}
\label{figmjrat}
\end{figure}

Although the general Jeans equation in spherical symmetry can be written
\[
{\partial \left (\nu \overline{v_r} \right ) \over \partial t} 
+ {\partial \left (\nu \overline{v_r^2}\right) \over \partial r}
+ {\nu \over r} \,\left [2\,\overline{v_r^2} - 
\left (\overline{v_\theta^2} + \overline{v_\phi^2} \right )\right ] + {\nu
\cot\theta\over r}\,\overline{v_rv_\theta} = - \nu\,{\partial \Phi\over
\partial r}
\ ,
\]
when kinematical modelers of observations write the Jeans equation, they
always omit the terms in $\overline {v_r v_\theta}$ and the term
with a partial time-derivative. 
With these omissions, the derived mass in a shell is

\begin{equation}\label{mjeans}
M_{\rm Jeans}(r) = - {r\left\langle v_r^2\right\rangle\over G}\,
\left (
{d\ln\rho\over d\ln r}
+
{d\ln\left\langle v_r^2\right\rangle \over d\ln r}
+ 2\,\beta
\right ) \ .
\end{equation}
Figure~\ref{figmjrat} plots the ratio of $M_{\rm Jeans}(r)/M(r)$.
One sees that the Jeans mass is equal to the true mass within a few percent
out to roughly $8\,R_e$. This means that the system is close enough to being
stationary that one can safely omit the time-derivative in the Jeans
equation. Beyond $8\,R_e$, the crossing time is long and the system has not
had time to relax to a stationary state.

\subsection{How much dark matter from orbital modeling?}

\begin{figure}[ht]
\centering \includegraphics[width=8.4cm]{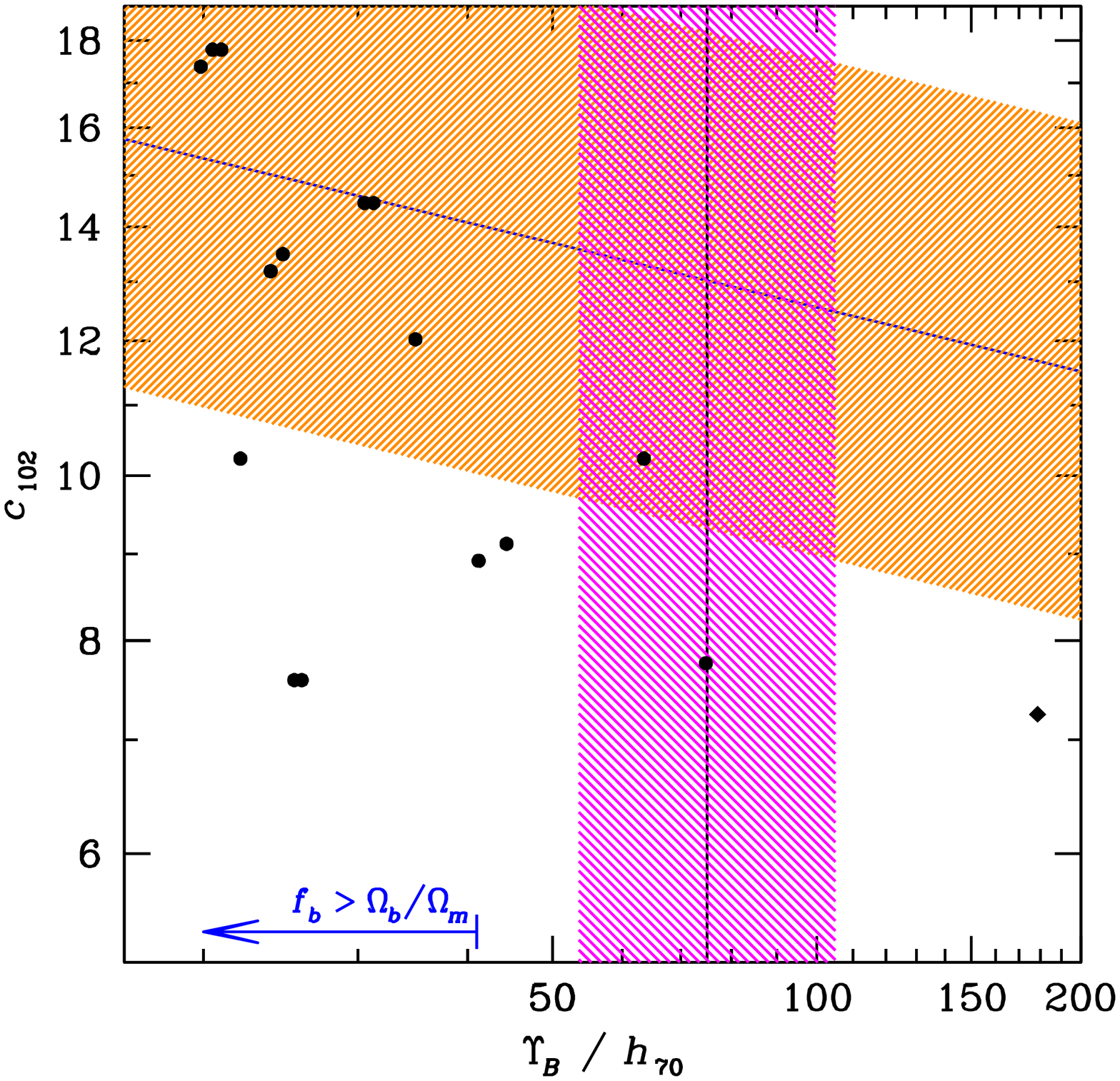}
\caption{Concentration vs. mass to blue light ratio within the virial radius
($r_{102}$) for
the orbit solutions for the elliptical galaxy NGC 3379
of Romanowsky et al. (2003) (\emph{circles}) and
an additional solution later found by A. Romanowsky (2005, private
communication, \emph{diamond}), adapted from Mamon \&
{\L}okas (2005b). 
For NGC 3379, we assume an absolute
magnitude $M_B = -20.12$, 
derived from a S\'ersic fit to the $B$-band surface brightness
profile of  de Vaucouleurs \& Capaccioli (1979), and corrected for galactic
extinction.
The \emph{vertical shaded region} is the recent $\Lambda$CDM constraint on $M/L$
within the virial radius by Eke et al. (2005).
The \emph{oblique shaded region} is the constraint on the $\Lambda$CDM
dark matter
concentration mass relation of Bullock et al. (2001), as rederived by
Napolitano et al. (2005) for $\sigma_8 = 0.9$. For both relations, we assume 
a 40\% uncertainty.}
\label{figcUpsRom}
\end{figure}

\nocite{Eke+05}
\nocite{dVC79}
\nocite{Napolitano+05}
While 
\cite{Romanowsky+03}
announced that their orbit modeling of NGC
3379 yielded no solutions with appreciable amounts of dark matter, a closer
inspection of their 15 orbit solutions (Fig.~\ref{figcUpsRom}),
shows that 2 solutions have the expected relatively high $M/L_B$ at the virial
radius, one of which has the expected dark matter concentration, while an
additional solution later found by Romanowsky (priv. comm.) has an 
even higher $M/L_B$. Therefore, \emph{the kinematic data (stars+PN) analyzed by
\citeauthor{Romanowsky+03}
is in fact consistent with the $\Lambda$CDM predictions}! 



\setlength\bibsep{2pt}
\newcommand{\bibfont}{\small}
\bibliography{master}
\end{document}